# Heteronuclear Polarization Transfers Between Spin-locked and Anti-Longitudinal Spin States in the NMR of Liquids and Spinning Solids

Sundaresan Jayanthi,* Adonis Lupulescu,* Julia Grinshtein and Lucio Frydman*

Department of Chemical and Biological Physics, Weizmann Institute of Science, Rehovot, Israel

## Abstract

Recently, Pang *et al* reported a novel polarization transfer scheme applicable to three-spin systems, whereby a rotating-frame NMR analogue of the cross effect could transfer polarization between; e.g.., two $^{13}$Cs and an $^{15}$N in a single crystal. The present work furthers this scheme to the case of powder NMR under magic angle spinning (MAS) conditions, as well as to solution NMR. It is found that in all such cases a second-order average Hamiltonian can transfer polarization between non-equivalent, coupled abundant spins (e.g., two $^1$Hs) prepared in "anti-longitudinal" magnetization states, and the spin-locked magnetization of a rare spins (e.g., one $^{13}$C). The average Hamiltonian for such three-spin $(S_1\text{-}S_2)\rightarrow I$ transfer was derived for both liquids and solids, and found in good quantitative agreement with numerical simulations and experiments. At an optimal transfer condition whereby an *I*-spin RF irradiation field matches the $S_1$-$S_2$ chemical-shift-difference, a maximum polarization enhancement of $\gamma_S/\gamma_I$ is achieved; as explained and demonstrated in the study, ca. half of this can be effectively obtained for $I = {}^{13}$C in powdered solids and in multi-spin systems in solutions. All such processes display an oscillatory nature, meaning that the transverse spin-locked polarization of a rare spin can become 'anti-longitudinal' $\pm(S_{1z} - S_{2z})$ magnetization of abundant spins –without ever pulsing on the latter. The roles played by many-body interactions, RF inhomogeneities, and interferences of other coherences during the execution of these novel forms of cross-polarization were investigated, and are exemplified with experiments and simulations.

**Emails**: jayanthi.sundaresan@weizmann.ac.il, Adonis.lupulescu@weizmann.ac.il, lucio.frydman@weizman.ac.il

# 1) Introduction

Nuclear magnetic resonance (NMR) relies on phenomena like the Overhauser Effect or on double-resonance pulsed schemes like INEPT or Hartmann-Hahn cross-polarization (HH CP), for transferring magnetization between homo- and hetero-nuclear spin systems.[1-5] Unidirectional polarization transfer schemes from sensitive spins $S$ (e.g., $^{1}$Hs) to insensitive heteronuclei $I$ (e.g., $^{13}$C) are essential for enhancing the latter spectra in directly detected 1D NMR experiments, whereas reciprocal transfers from abundant-to-heteronuclei and back, form essential components of modern multidimensional NMR correlations.[6,7] Both INEPT and HH CP can proceed between single spin pairs, usually mediated by heteronuclear J- or dipolar-couplings in liquids and solids, respectively. CP is also possible between the dipolar order that in the solid state can arise when dealing with multiple abundant $S$-spins (e.g., $^{1}$Hs); $I$-nuclei irradiated at ca. the homonuclear dipolar coupling strength within such ensemble can then transform such order into observable polarization.[8,9] Two dipole-coupled $S$-spins in static solids and three mutually coupled $S$-spins in solids subject to magic-angle-spinning (MAS) are minimal ensembles capable of supporting such homonuclear-dipolar→heteronuclear-magnetization CPs,[10-13] which have no correspondence in solution-state experiments. Recently, Pang *et al* reported an additional CP scheme that may also arise in three-spin systems, termed as the nuclear cross effect.[14] In parallel with the cross-effect arising in dynamic nuclear polarization (DNP),[15-19] this transfer of polarization requires two spins $S_1$, $S_2$ whose difference in Larmor frequency matches a resonance frequency of the receiving species $I$, as well as an imbalance between the two $S$-spin polarizations. By contrast to the DNP cross-effect, however, the $I$ frequency that matters in this NMR cross effect is not the Larmor one, but rather the Rabi frequency that is imparted on $I$ by a spin-locking field: the transfer thus happens in the rotating, rather than in the laboratory frame.

The present study revisits such scenario, considering again transfers between abundant spins $S_1$, $S_2$ (e.g., two inequivalent $^{1}$Hs) and an insensitive spin $I$ (e.g., a $^{13}$C). It is shown that $S$-spins prepared in an "anti-longitudinal' state $\pm(S_{1z} - S_{2z})$ –or in general, in an out-of-equilibrium state– can then be converted into $^{13}$C magnetization via a CP that mediated by either dipolar or J-couplings. These transfers are shown to occur in both liquids and spinning solids, provided that (i) the $I$-spin is spin-locked in a transverse state with an RF field that matches the chemical shift difference between the coupled $S$-spins, and (ii) the $I$-$S_1$ and $I$-$S_2$ couplings driving this three-spin process, are unequal (Figure 1a). Conversely, a transverse $I$-spin magnetization that is spin-locked with an RF field matching the chemical shift difference between coupled *S*-spins, can generate an anti-longitudinal $S_{1z} - S_{2z}$ spin state –without ever having to pulse on the latter species (Figure 1b). Enabling in all cases these heteronuclear polarization transfers are flip-flop terms in the homonuclear *S*-spin dipolar or J-coupling Hamiltonians, which though truncated in liquids by high-fields and in solids by MAS, suffice to drive these dynamics. A theory covering the ensuing multi-spin CP schemes in liquids and spinning solids is provided, including average Hamiltonians that are in quantitative agreement with brute-force numerical simulations. Also accompanying these theories are MAS and liquid-state experiments verifying the theoretical predictions. The roles of many-body interactions and rf-inhomogeneity in the transfer process, as well as differences and similarities with conventional Hartmann-Hahn based and dipole-order-based CP, are briefly discussed.

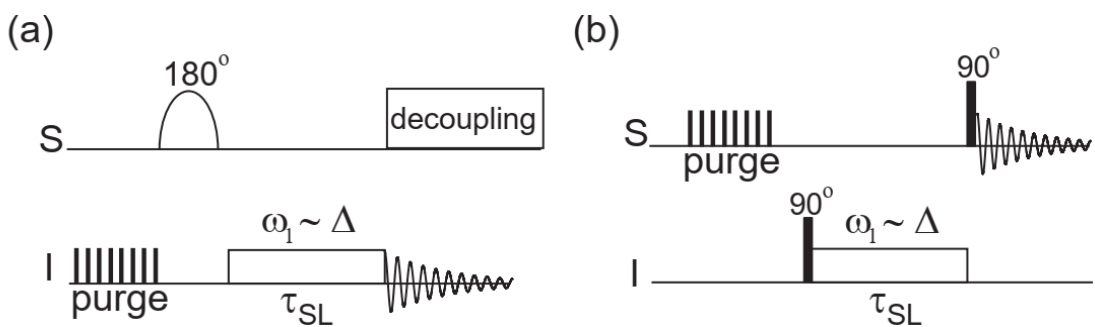

Figure 1: Pulse scheme employed for (a) polarization transfer from anti-longitudinal *S*-spin magnetization to the rare *I* spin, followed by observation under heteronuclear decoupling. (b) Creation and detection of anti-longitudinal *S*-spin magnetization from rare spins. $\tau_{SL}$ denotes the spin-lock time and $\omega_1$ the RF field applied for the transfer, which under optimal 3-spin CP conditions is approximately the chemical shift difference *Δ*, (in angular frequency units) between the *S*-spins involved in the anti-longitudinal state. This scheme can be implemented for both liquids and spinning solids. In the actual experiments a series of high-power "purging" pulses of alternating phases were used to delete prior polarization; in liquid experiments gradients were incorporated in-between the purge pulses and after the selective pulse. In (b), the $^{13}$C 90° pulse along with the receiver were phase cycled as ±x.

## 2) Theoretical Description

This section provides a unified theoretical description of the 3-spin CP scheme discussed in this paper. The approach was developed based on Average Hamiltonian Theory (AHT),[20,21] and although higher order terms of the Magnus expansion had to be used to achieve convergence, it provides valuable insight on the spin dynamics of the transfer for both liquids and spinning solids. For simplicity our analyses focused on an anti-longitudinal magnetization $(S_{1z} - S_{2z})$ of two inequivalent, mutually-coupled abundant spins $S_1$, $S_2$ as starting point, that are in turn coupled to a third heteronucleus (*I*).[1] In all cases the basic frequency that was considered for deriving the relevant average Hamiltonians, was the chemical shift difference between $S_1$ and $S_2$.

## 2.a Evaluating the Three-Spin CP Average Hamiltonian in Liquids

We consider two spin-1/2 nuclei, $S_1$ and $S_2$, *J*-coupled among themselves and *J*-coupled to a third hetero-spin *I*. Under on-resonance irradiation on the *I*-spins, the ensuing rotating frame Hamiltonian is[6,7]

$$\mathcal{H} = \omega_1 I_x + \Omega_1 S_{1z} + \Omega_2 S_{2z} + \omega_J^{SS}[S_{1z}S_{sz} + 1/2(S_{1+}S_{2-} + S_{1-}S_{2+})] + \omega_J^{S1I} S_{1z} I_z \\ + \omega_J^{S2I} S_{2z} I_z \qquad (1)$$

where $\omega_J^{SS}, \omega_J^{S1I}, \omega_J^{S2I}$ are the homonuclear and heteronuclear *J*-couplings, $\Omega_1, \Omega_2$ are the respective *S*-spin chemical shifts, and $\omega_1$ is the amplitude of the RF field applied on the *I*-spin. For consistency, the strengths of all these couplings, shifts and fields are expressed in angular frequency units. It is convenient to re-express the above Hamiltonian in a tilted *I*-spin frame whose z-axis is defined by the RF irradiation, according to the transformation $\mathcal{H} = \exp\left(-i\frac{\pi}{2} I_y\right) \mathcal{H}_T \exp\left(+i\frac{\pi}{2} I_y\right)$. This leads to

$$\mathcal{H}_T = \omega_1 I_z + \Omega_1 S_{1z} + \Omega_2 S_{2z} + \omega_J^{SS}[S_{1z}S_{sz} + 1/2(S_{1+}S_{2-} + S_{1-}S_{2+})] - \omega_J^{S1I} S_{1z} I_x \\ - \omega_J^{S2I} S_{2z} I_x \qquad (2)$$

Denoting the basis states of the $S_1$-$S_2$ subspace as $|1\rangle = |\alpha\alpha\rangle, |2\rangle = |\alpha\beta\rangle, |3\rangle = |\beta\alpha\rangle, |4\rangle = |\beta\beta\rangle$, $\mathcal{H}_T$ can be represented in terms of fictitious $S_1$, $S_2$ and *I* spin-half operators as

$$\mathcal{H}_T = \omega_1(\mathbf{1}^{1-4} + \mathbf{1}^{2-3})I_z + \Sigma\, S_z^{1-4} + \Delta\, S_z^{2-3} + \frac{\omega_J^{SS}}{4}(\mathbf{1}^{1-4} - \mathbf{1}^{2-3}) + \omega_J^{SS} S_x^{2-3} - J_+ S_z^{1-4} I_x \\ - J_- S_z^{2-3} I_x, \qquad (3)$$

where $\Sigma = \Omega_1 + \Omega_2; \Delta = \Omega_1 - \Omega_2 > 0; J_+ = \omega_J^{S1I} + \omega_J^{S2I}; J_- = \omega_J^{S1I} - \omega_J^{S2I}$; and $\mathbf{1}^{1-4}$ and $\mathbf{1}^{2-3}$ are unit operators of the respective double- and zero-quantum *S* subspaces. It can be seen that $\mathcal{H}_T$ is a sum of two commuting terms: $\mathcal{H}_T^{1-4}$ and $\mathcal{H}_T^{2-3}$.

---

[1] Should the initial state involve other unequal $S_{1z}$, $S_{2z}$ populations (e.g. an $S_{2z}$ saturation), these can always be expressed as $\rho_0 = \alpha(S_{1z}+S_{2z}) + \beta(S_{1z}-S_{2z})$. Since as will be shown below ($S_{1z}+S_{2z}$) commutes with the relevant Hamiltonian and becomes immaterial, an analysis based solely on an ($S_{1z}$-$S_{2z}$) initial state, suffices for all cases.

For an initial state $\rho(0) = S_{1z} - S_{2z} = 2S_z^{2-3}$ which is confined to the 2-3 subspace, the terms $\frac{\omega_J^{SS}}{2} \mathbf{1}^{1-4}$ and $\frac{\omega_J^{SS}}{2} \mathbf{1}^{2-3}$ may be dropped, as they are energy shifts that do not contribute to evolution. The relevant sub-space Hamiltonians are thus

$$\mathcal{H}_T^{1-4} = \omega_1 \mathbf{1}^{1-4} I_z + \Sigma S_z^{1-4} - J_+ S_z^{1-4} I_x, \tag{4a}$$

$$\mathcal{H}_T^{2-3} = \omega_1 \mathbf{1}^{2-3} I_z + \Delta S_z^{2-3} + \omega_J^{SS} S_x^{2-3} - J_- S_z^{2-3} I_x. \tag{4b}$$

Since $[\mathcal{H}_T^{1-4}, \mathcal{H}_T^{2-3}] = 0$, the total propagator $U_T(t,0)$ for this three-spin system factorizes into $U_T(t,0) = U_T^{1-4}(t,0) U_T^{2-3}(t,0)$; further, since $[\mathcal{H}_T^{1-4}, \rho(0)] = 0$, the spins' time evolution will be guided solely by $U_T^{2-3}(t,0)$:

$$\rho(t) = U_T^{2-3}(t,0) \rho(0) U_T^{2-3}(t,0)^+. \tag{5}$$

Transforming to an interaction frame defined by

$$U_T^{2-3}(t,0) = \exp[-i\Delta(S_z^{2-3} + I_z)t] \tilde{U}_T^{2-3}(t,0), \tag{6}$$

the Hamiltonian $\tilde{\mathcal{H}}_T^{2-3}$ corresponding to $\tilde{U}_T^{2-3}$ will be given by

$$\tilde{\mathcal{H}}^{2-3}(t) = (\omega_1 - \Delta) I_z + \omega_J^{SS} \left[ S_x^{2-3} \cos(\Delta t) - S_y^{2-3} \sin(\Delta t) \right]$$
$$- J_- S_z^{2-3} \left[ I_x \cos(\Delta t) - I_y \sin(\Delta t) \right]. \tag{7}$$

$\tilde{\mathcal{H}}^{2-3}(t)$ displays oscillatory behaviour at frequency $\Delta$. Therefore, an effective Hamiltonian can be obtained for it by applying AHT with a basic period $1/\Delta$. In order to do this, we rewrite the Hamiltonian $\tilde{\mathcal{H}}^{2-3}$ as

$$\tilde{\mathcal{H}}^{2-3}(t) = \sum_p \tilde{H}_p \exp(ip\Delta t),$$

where only three Fourier components, $\tilde{H}_0$ and $\tilde{H}_{\pm 1}$ are present:

$$\tilde{H}_0 = (\omega_1 - \Delta) I_z, \quad \tilde{H}_{+1} = \frac{1}{2} \omega_J^{SS} S_+^{2-3} - \frac{1}{2} J_- S_z^{2-3} I_+, \quad \tilde{H}_{-1} = \frac{1}{2} \omega_J^{SS} S_-^{2-3} - \frac{1}{2} J_- S_z^{2-3} I_- \tag{8}$$

The first-order average Hamiltonian is $\mathcal{H}_{ave}^{(1)} = \tilde{H}_0$, and leads to no CP transfer. The second-order average Hamiltonian can be computed from the Fourier terms of $\tilde{H}_p$ (see Supporting Information (SI), Section I). It is then found that $\mathcal{H}_{ave}$ up to second order is

$$\mathcal{H}_{ave} = (\omega_1 - \Delta) I_z + A \left( I_z + S_z^{2-3} \right) + B \left( I_z - S_z^{2-3} \right) + C \, ZQ_x + D S_z^{2-3} I_x \tag{9}$$

where $ZQ_x = (I_+S_-^{2-3} + I_-S_+^{2-3})/2$, and $A = \left(J_-^2 + \left(2\omega_J^{SS}\right)^2\right)/16\Delta$, $B = \left(J_-^2 - \left(2\omega_J^{SS}\right)^2\right)/16\Delta$, $C = \left(J_-\omega_J^{SS}\right)/2\Delta$. We shall neglect the last term $D = -(\omega_1 - \Delta)J_-/\Delta$, as transfer conditions will have to meet $\omega_1 \approx \Delta$ and therefore it is safe to assume that $\omega_1 - \Delta \ll \Delta$.[2]

The Hamiltonian of Eq. (9) can be further simplified by representing it as

$$\mathcal{H}_{ave} = \frac{(A+B)}{2}(I_z + S_z^{2-3}) + \frac{(A-B)}{2}(I_z - S_z^{2-3}) + CZQ_x \tag{10}$$

The first term commutes with the rest of the terms as well as with the initial density operator $\rho(0) = 2S_z^{2-3}$; therefore, it is sufficient to consider the simpler form

$$\mathcal{H}_{ave} \approx \frac{(A-B)}{2}(I_z - S_z^{2-3}) + C\,ZQ_x \tag{11}$$

Equation (11) resembles the $(\omega_1^I - \omega_1^S)(I_z - S_z) + \omega_J^{IS} ZQ_x$ HH CP Hamiltonian in the zero-quantum subspace of an I-S interaction frame.[22,23] Hence, an analogous matching condition calls for nulling the coefficient of the $(I_z - S_z^{2-3})$ term. For the 3-spin CP case this leads to the RF matching condition

$$\omega_1^{opt} = \Delta + \frac{\left(\omega_J^{SS}\right)^2}{2\Delta} - \frac{(J_-)^2}{8\Delta}, \tag{12}$$

which requires the I-spin RF to match the difference in S-spin chemical shifts, after correcting for homonuclear S-spin coupling effects and for differences in the heteronuclear J-coupling effects. Setting the RF to such $\omega_1^{opt}$ value results in the average Hamiltonian

$$\mathcal{H}_{ave} = CZQ_x. \tag{13}$$

Notice that, to first order, this matching condition is similar to that presented in Eq. (6) of reference [15].

Assuming that the I-spin RF has been set to this $\omega_1^{opt}$ condition, the state of the system after an irradiation time $\tau$ can be rewritten in the tilted frame as

$$\rho(\tau) = \frac{1}{2}(1 - \cos(C\tau))I_z + \frac{1}{2}(1 + \cos(C\tau))(S_{1z} - S_{2z}) - 2(1 - \cos(C\tau))I_zS_{1z}S_{2z}$$
$$- 2\sin(C\tau)\frac{[I_-S_{1+}S_{2-} - I_+S_{1-}S_{2+}]}{2i}. \tag{14}$$

---

[2]Notice also that $ZQ_x$ resembles a two-spin zero-quantum operator; however, due to the presence of $I_+$ and $I_-$, it is in fact a single-quantum operator (strictly speaking the $S_\pm^{2-3}$ being zero-quantum operators).

After passing back from the tilted to the usual rotating frame, the density operator becomes

$$\rho(\tau) = \frac{1}{2}(1 - \cos(C\tau))I_x + \frac{1}{2}(1 + \cos(C\tau))(S_{1z} - S_{2z}) - 2(1 - \cos(C\tau))I_x S_{1z} S_{2z}$$
$$- 2\sin(C\tau)\frac{[(-I_z - iI_y)S_{1+}S_{2-} - (-I_z + iI_y)S_{1-}S_{2+}]}{2i} \quad (15)$$

Notice the similarities between the three initial terms in this expression, and those arising in the HH CP case, with now $(S_{1z} - S_{2z})$ taking the place of the spin-locked $S_x$ magnetization from which polarization originates. If, following an RF-driven transfer executed under optimal matching conditions, an $I$-spin (e.g. $^{13}$C) acquisition is performed in the presence of $S$-spin (e.g. $^1$H) broadband decoupling, only the first term in Eq. (15) will contribute to the detected signal.

For simplicity, the analysis above only considered the spin operators. Taking into account the appropriate differences in gyromagnetic ratios and repeating the analysis leading to Eq. (15), it follows that an $I$-spin polarization enhancement of magnitude $\gamma_S/\gamma_I$ can be achieved by the 3-spin CP here discussed. For a $^1$H-$^{13}$C spin system this amounts to a possible 4-fold enhancement of the $^{13}$C signal –the same as in HH CP or INEPT. In such traditional polarization transfer schemes, the optimum $S \rightarrow I$ transfer buildup time is $\propto 1/\omega_J^{IS}$;[3,4,22-24] by contrast, for the three-spin case here considered, Eq. (15) predicts a rate $C = (J_-\omega_J^{SS})/2\Delta$. $J_- = \omega_J^{S1I} - \omega_J^{S2I}$ takes here the role of the heteronuclear coupling strength $\omega_J^{IS}$; however, this is now scaled by a factor $\propto \omega_J^{SS}/\Delta$ –i.e., by the extent to which the homonuclear $S$-spins deviate from the weak coupling regime. This reflects the second-order nature of this 3-spin CP process, associated to longer buildup times than HH CP, and to an inverse static magnetic field ($B_0$) dependence.

Figures 2a-2c illustrate the accuracy of the theory just presented (dashed lines), when compared with brute-force numerical simulations (red), for a simple three-spin system. Calculations are presented as a function of $\Delta$, for an ensemble subject to the sequence in Figure 1a. Notice the slight discrepancies between the predictions of Eqs. (13)-(15) and exact simulations at lower $\Delta$ values; this is as expected, given the slower convergence that the Magnus expansion will have in such cases. For higher chemical shift $\Delta$ values the agreement between second-order AHT and brute-force simulations is restored; in such cases, both relaxation-free models predict a full $\gamma_S/\gamma_I$ enhancement of the $I$-spin polarization, happening in synchrony with oscillations of the anti-longitudinal $S$-spin polarization. Figures 2d-2f consider the dependence of this CP transfer on the RF amplitude. A matching condition around

the optimal value $\omega_1^{opt}$ given in Eq. (12) is clearly seen, which approaches the exact *S*-spin chemical shift difference as Δ becomes larger. When $\omega_1$ does not satisfy this optimal matching condition the presence of the $(I_z - S_z^{2-3})$ term in Eq. (11) rapidly quenches the polarization transfer. This leads to a sharp matching condition, and suggests that RF inhomogeneity may influence the efficiency of the transfer; this is indeed verified by the simulations in Figures 2g-2i. Even modest 5% RF inhomogeneities like the ones present in high-resolution NMR probes (blue plots), can substantially reduce the achieved polarization transfer. A possible remedy to this drawback consists of utilizing linearly ramped RF fields to implement the S→I transfer; this strategy works fairly well (Figs. 2g-2i, red plots) at the cost of increased irradiation times.

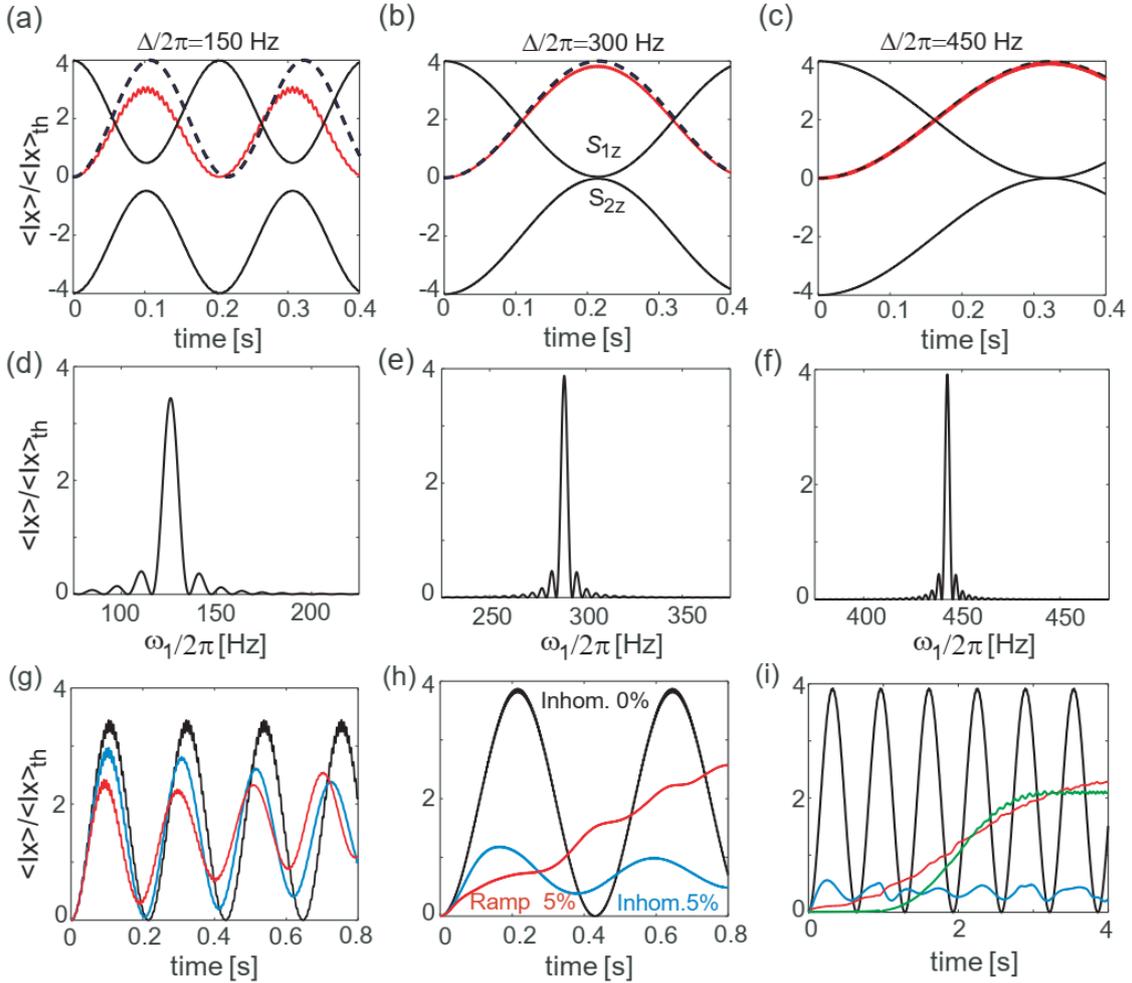

Figure 2. (a-c) CP transfer processes predicted by AHT (black lines; dashed ones correspond to $I_x$) versus brute-force numerical simulations (red), for $\Delta/2\pi$ = 150, 300, and 450 Hz. The evolution of $\langle S_{1z}\rangle/\langle I_z\rangle_{th}, \langle S_{2z}\rangle/\langle I_z\rangle_{th}$ (black) is also shown, where $\langle I_z\rangle_{th}$ is the thermal expectation value for *I*. An $H_1 - H_2 - {}^{13}C$ spin system was considered, giving $\langle S_{1z}\rangle_{th}/\langle I_z\rangle_{th} \sim 4, etc$. (d-f) RF matching profile of the polarization transfer displaying high selectivity. The optimum RF amplitudes, $\omega_1^{opt}/2\pi =$ 127.8, 288.6 $and$ 442.4 $Hz$, agree very well with the predictions of Eq. (12). (g-i) Polarization transfer for an RF inhomogeneity of 0% (black) and 5% (color), with constant RF amplitude and a linear ramp of 5% in blue and red respectively. In (i) a linear ramp of 10% (green, 5% RF inhomogeneity) is also shown.

Other parameters are $\omega_J^{HH}/2\pi = 8.5$ Hz, $\omega_J^{H1C}/2\pi = 172$ Hz, $\omega_J^{H2C}/2\pi = 8$ Hz. In all cases the $^{13}$C offset was set to zero.

These treatments and calculations considered an isolated $S_1S_2I$ system. However, for most molecules, larger spin systems may have to be considered: protons rarely come in isolated pairs. Although we have not derived a clear rule of what will be the effects of considering such multi-spin systems, we found (Supporting Figure S1) that the presence of more than two $S$-spins tends to decrease the transfer efficiency. These and other effects are further treated in the SI, Section I 1.

Continuing with the quantum-mechanical treatment introduced above, one can also prove that an initial $I_x$ polarization, spin-locked and subject to RF irradiation at an appropriate matching condition (Figure 1b), will originate an anti-longitudinal $S_{1z} - S_{2z}$ spin state – *without ever having to pulse on the S-spins*. This "inverse polarization" process starts in the tilted frame description with $\rho(0) = I_z$, and evolves upon suitable matching as

$$\rho(\tau) = \left[\frac{3}{4} + \frac{1}{4}\cos(C\tau)\right]I_z + \frac{1}{4}[1 - \cos(C\tau)](S_{1z} - S_{2z}) + [1 - \cos(C\tau)]I_zS_{1z}S_{2z}$$
$$+ \sin(C\tau)\frac{[I_-S_{1+}S_{2-} - I_+S_{1-}S_{2+}]}{2i}. \quad (16)$$

To make such transferred anti-magnetization state visible, a 90° readout S-pulse would have to be applied. After passing back from the tilted frame and applying such 90° S-pulse at a time $\tau$, the relevant signal-generating states during detection will be

$$\rho(\tau) = \frac{1}{4}[1 - \cos(C\tau)](S_{1x} - S_{2x}) - \sin(C\tau)\frac{I_z[(S_{1z}S_{2y} - S_{1y}S_{2z})]}{2}. \quad (17)$$

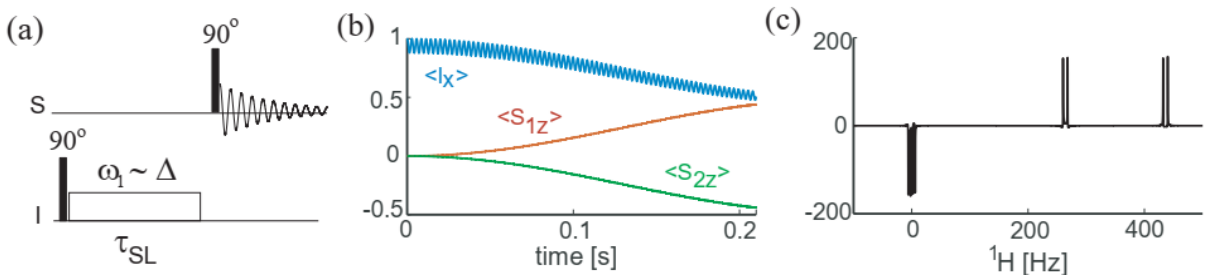

Figure 3. Creation of anti-longitudinal $S$-spin polarization from an initial $I$-spin transverse magnetization, subject to spin-locking irradiation close to the $\omega_1 \approx \Delta$ matching condition. A three-spin ($H1, H2, C$) system was here considered, with the following parameters: $\Delta/2\pi = 350$ Hz, $J_{H1C} = 172$ Hz, $J_{H2C} = 4$ Hz, $J_{H1H2} = 8$ Hz, $\omega_1/2\pi = 340$ Hz. (a) Simulated pulse sequence. (b) Time buildup terminated at the maximum of $\langle S_{1z}\rangle - \langle S_{2z}\rangle$, which is 216.3 ms. The quantities being plotted are $\langle I_x\rangle/\langle I_z\rangle_{th}$, $\langle S_{1z}\rangle/\langle I_z\rangle_{th}$, and $\langle S_{2z}\rangle/\langle I_z\rangle_{th}$. The fast oscillations exhibited by $\langle I_x\rangle$ are at frequency $\Delta$. (c) Corresponding $^1$H spectra arising after a 90° proton pulse.

The first of these terms is the anti-longitudinal state being sought. Without decoupling on the I-channel, the second term will also lead to an antiphase structure in the *S*-spectrum, as driven by the homonuclear J-coupling. This would contaminate the expected pureness of the anti-longitudinal spectral contribution, but its contribution can be eliminated by an additional 90°(y) pulse on I-channel. Alternatively, decoupling on the *I*-channel eliminates this second term automatically. Elimination of this second term gives a maximal amplitude for the anti-longitudinal polarizations equal to $\gamma_I/2\gamma_S$ –clearly not a gain if pertaining *S*=proton polarization transfer experiments. Further features pertaining to the above discussion are shown in Figure 3.

**2.b Evaluating the Three-spin CP Average Hamiltonian in Spinning Solids**

As next step in this analysis we extend the three-spin CP formalism to solids. For the case of a static single crystal, and thanks to the analogous form taken by the interactions, an approach similar to that in Section 2.a can be used to derive the RF matching condition and other details of the transfer. One then arrives at $\omega_1^{opt} = \Delta + \frac{d_{SS}^2}{2\Delta} - \frac{(D_{S1I}-D_{S2I})^2}{2\Delta}$, where $d_{SS}, D_{S1I}, D_{S2I}$ are the homonuclear and heteronuclear dipolar couplings (for simplicity we disregard J-couplings), and $\Delta$ the $S_1$-$S_2$ shift difference as dictated by isotropic and anisotropic chemical shieldings. Further pursuing such static solid analysis, however, is of limited value. First, because in parallel to the liquids case the aforementioned matching condition will only be valid for small $d_{SS}^2/2\Delta$ and $(D_{S1I} - D_{S2I})^2/2\Delta$ ratios; this is a constrain which is rarely fulfilled in rigid solids, involving proton networks and/or directly-bonded proton-carbon spin pairs. Further, for solid samples, fast/ultrafast MAS rates might be indispensable to selectively invert/resolve otherwise strongly dipole-coupled proton resonances. It follows that considering this kind of experiments on static rigid solids makes little practical sense.

We therefore focus on fast MAS cases, such that proton resonances involved in the transfer are sufficiently resolved, homo- and heteronuclear dipolar couplings average over a rotational period to zero, and chemical shifts due likewise to their isotropic values. As in the liquid case we consider two spin-1/2, $S_1$ and $S_2$, coupled among themselves as well as to a third hetero-spin *I* –except that now all couplings are dipole-dipole. Assuming MAS at an angular frequency $\omega_R$ and an on-resonance *I*-spin irradiation with an RF amplitude $\omega_1$, the relevant Hamiltonian will now be

$$\mathcal{H}(t) = \omega_1 I_x + \Omega_1 S_{1z} + \Omega_2 S_{2z} + d_{SS}(t)[2S_{1z}S_{sz} - 1/2(S_{1+}S_{2-} + S_{1-}S_{2+})] + d_{S1I}(t)2S_{1z}I_z$$
$$+ d_{S2I}(t)2S_{2z}I_z \quad (18a)$$

where $d_{SS}(t), d_{S1I}(t), d_{S2I}(t)$ are the respective time-dependent dipolar couplings, and $\Omega_1, \Omega_2$ are chemical shifts of the $S$ spins that for simplicity we assume are time-independent, isotropic values. MAS will modulate the dipolar couplings as[21,22]

$$d_{SS}(t) = \sum_n d_{SS}^{(n)} e^{in\omega_R t}; \quad d_{S1I}(t) = \sum_n d_{S1I}^{(n)} e^{in\omega_R t}; \quad d_{S2I}(t) = \sum_n d_{S2I}^{(n)} e^{in\omega_R t} \quad (18b)$$

where the various $\{d^{(n)}\}$ Fourier coefficients are as defined in Reference [25]. Subjecting the Hamiltonian in Eq. (18) to a tilted frame transformation $\mathcal{H} = \exp\left(-i\frac{\pi}{2} I_y\right) \mathcal{H}_T \exp\left(+i\frac{\pi}{2} I_y\right)$, and representing the fictitious $S_1$-$S_2$ spin-half operators as before, we arrive to a tilted-frame Hamiltonian (see SI, Section II)

$$\mathcal{H}_T(t) = \omega_1(\mathbf{1}^{1-4} + \mathbf{1}^{2-3})I_z + \Sigma S_z^{1-4} + \Delta S_z^{2-3} + \frac{d_{SS}(t)}{2}(\mathbf{1}^{1-4} - \mathbf{1}^{2-3}) - d_{SS}(t) S_x^{2-3}$$
$$- 2D_+(t)S_z^{1-4}I_x - 2D_-(t)S_z^{2-3}I_x. \quad (19a)$$

where

$$\Sigma = \Omega_1 + \Omega_2; \Delta = \Omega_1 - \Omega_2 > 0; D_+(t) = d_{S1I}(t) + d_{S2I}(t); D_-(t) = d_{S1I}(t) - d_{S2I}(t) \quad (19b)$$

As before, this $\mathcal{H}_T(t)$ is a sum of two commuting terms $\mathcal{H}_T^{1-4}(t)$ and $\mathcal{H}_T^{2-3}(t)$

$$\mathcal{H}_T^{1-4}(t) = \frac{d_{SS}(t)}{2}\mathbf{1}^{1-4} + \omega_1 \mathbf{1}^{1-4}I_z + \Sigma S_z^{1-4} - 2D_+(t)S_z^{1-4}I_x \quad (20a)$$

$$\mathcal{H}_T^{2-3}(t) = -\frac{d_{SS}(t)}{2}\mathbf{1}^{2-3} + \omega_1 \mathbf{1}^{2-3}I_z + \Delta S_z^{2-3} - d_{SS}(t) S_x^{2-3} - 2D_-(t)S_z^{2-3}I_x \quad (20b)$$

Assuming as in the liquid case an initial "anti-magnetization" state $\rho(0) = S_{1z} - S_{2z} = 2S_z^{2-3}$ confined to the 2-3 subspace, both the $\mathcal{H}_T^{1-4}(t)$ and the term $-\frac{d_{SS}(t)}{2}\mathbf{1}^{2-3}$ can be omitted from further consideration. The relevant propagation is thus solely

$$\rho(t) = U_T^{2-3}(t, 0)\rho(0)U_T^{2-3}(t, 0)^+ \quad (21)$$

which can be further simplified by transforming to an interaction frame given by

$$U_T^{2-3}(t, 0) = \exp[-i\Delta(S_z^{2-3} + I_z)t] \widetilde{U}_T^{2-3}(t, 0). \quad (22)$$

where the Hamiltonian $\widetilde{H}_T^{2-3}$ corresponding to $\widetilde{U}_T^{2-3}$ is

$$\widetilde{\mathcal{H}}^{2-3}(t) = (\omega_1 - \Delta)I_z - d_{SS}(t)\left[S_x^{2-3}\cos(\Delta t) - S_y^{2-3}\sin(\Delta t)\right]$$
$$- 2D_-(t)S_z^{2-3}\left[I_x\cos(\Delta t) - I_y\sin(\Delta t)\right]. \qquad (23)$$

So far this is as in the liquid state case; however, by contrast to Eq. (7), Eq. (23) has two different time dependencies: a slower one associated to $\Delta$, and a faster one in $D_-(t)$ and $d_{SS}(t)$ associated to $\omega_R$. In a general case, to first order, the average Hamiltonian $\widetilde{\mathcal{H}}^{2-3}_{(1)} = (\omega_1 - \Delta)I_z$; this is the only one surviving term, and it will quench any transfer process unless set to zero. This requires setting $\omega_1 = \Delta$, which constitutes the matching condition. To apply AHT we now set a simplifying $\omega_R = k\Delta$ assumption, with $k$ an integer; together with the $\omega_1 = \Delta$ matching this enables calculation of the second order average Hamiltonian $\widetilde{\mathcal{H}}^{2-3}_{(ave)}$ that ends up driving the polarization transfer. To do so we rewrite the $\widetilde{\mathcal{H}}^{2-3}(t)$ Hamiltonian as (see SI Section II for further details)

$$\widetilde{\mathcal{H}}^{2-3}(t) = \sum_p \widetilde{H}_p \exp(ip\Delta t) \ . \qquad (24)$$

where the non-vanishing Fourier components $\widetilde{H}_p$, have $p = nk \pm 1$ and $n = \pm 1, \pm 2$. For example, for $n = 1$, $\widetilde{H}_{k+1} = -\frac{1}{2}d_{SS}^{(1)}S_+^{2-3} - D_-^{(1)}S_z^{2-3}I_+$ and $\widetilde{H}_{k-1} = -\frac{1}{2}d_{SS}^{(1)}S_-^{2-3} - D_-^{(1)}S_z^{2-3}I_-$.

We use this expression to consider the ensuing average Hamiltonian over the long period $1/\Delta$, so that $n\omega_R$ in Eq. (18) becomes $nk\Delta$. Since to first order the Hamiltonian $\mathcal{H}^{(1)}_{ave} = \widetilde{H}_0$ is zero, we focus on the second order effects. These are

$$\mathcal{H}^{(2)}_{ave} = A(I_z + S_z^{2-3}) + B(I_z - S_z^{2-3}) + C_1 ZQ_x + C_2 ZQ_y, \qquad (25)$$

where the operators $ZQ_x = (I_+S_-^{2-3} + I_-S_+^{2-3})/2$, $ZQ_y = (I_+S_-^{2-3} - I_-S_+^{2-3})/2i$, represent quadrature 'zero-quantum' operators, and the explicit forms of the $A, B, C_1, C_2$ coefficients are given in the SI, Section II. All of these coefficients entail cross terms within or between the $d_{SS}^{(n)}$ and $D_-^{(n')}$ coefficients so that $n + n' = 0$, as well as an inverse dependence on $\Delta$ reflective of their higher-order nature.

As before, the average Hamiltonian in Eq. (25) is the sum of two terms: $\mathcal{H}^{(+)}_{ave} = A(I_z + S_z^{2-3})$ that commutes with the initial state $\rho(0) = 2S_z^{2-3}$ and hence does not matter, and $\mathcal{H}^{(-)}_{ave} = B(I_z - S_z^{2-3}) + C_1 ZQ_x + C_2 ZQ_y$ which does not commute and hence contributes to the spin dynamics. $\mathcal{H}^{(-)}_{ave}$ can be expressed as $\mathcal{H}^{(-)}_{ave} = B(I_z - S_z^{2-3}) + C_{eff}(cos(\chi)ZQ_x +$ $sin(\chi)ZQ_y)$, where $C_{eff} = \sqrt{C_1^2 + C_2^2}$ and $\chi = atan(C_2/C_1)$. In a suitably tilted frame within the $|3\alpha\rangle = |+\rangle, |2\beta\rangle = |-\rangle$ subspace this can be written as

$$\mathcal{H}_{ave}^{(-)} = 2BZ + C_{eff}X \qquad (26)$$

where $Z = [|+\rangle\langle+|-|-\rangle\langle-|]/2$, $X = [|+\rangle\langle-|+|-\rangle\langle+|]/2$. This is equivalent to the Hamiltonian of a spin-1/2 with an 'offset-like' term $2B$ and an 'RF-like' term $C_{eff}$. Eq. (26) can thus be further represented as,

$$\mathcal{H}_{ave}^B = \omega_{eff}[\cos\phi\, Z + \sin\phi\, X] \qquad (27)$$

where $\omega_{eff} = \sqrt{4B^2 + C_{eff}^2}$ and $\tan\phi = C_{eff}/2B$. The state of the system at time $t$ can then be readily evaluated; after returning from the tilted frame this becomes,

$$\rho(t) = \frac{1}{2}\sin^2\phi\left[1 - \cos(\omega_{eff}t)\right]I_x + [2 - \sin^2\phi\left(1 - \cos(\omega_{eff}t)\right)]S_Z^{2-3} + \text{other terms} \qquad (28)$$

where 'other terms' in Eq. (28) are much more complex than in liquids, but will likewise not contribute to the observable signal if the $I$-spin is detected with $S$-decoupling.

Figure 4 illustrates how the predictions of this analytical AHT (red) matches numerical brute-force simulations (black) for three-spins H1-H2-C arranged in an "L"-like configuration, placed at arbitrarily chosen orientations within a rotor and undergoing MAS. All calculations assume $\omega_R/2\pi = 40$ kHz and $\Delta/2\pi = 1$ kHz; i.e., $k = 40$. The correspondence between theory and simulation is excellent, showing no significant mismatches between the two. Agreement continues for values of $\omega_R/\Delta$ reaching down all the way to $k = 2$, where theories begin to diverge due to the onset of an alternative HORROR-like recoupling.[30] Notably, even if $\omega_R/\Delta$ does not match an integer $k$, the agreement between theory and simulation remains excellent. From these and other calculations (not shown; scripts available upon request) it can be concluded that, as in liquids, polarization transfers of up to $\gamma_S/\gamma_I$ relative to thermal I-spin polarization, can be obtained from an initial $S_{1z}$-$S_{2z}$ state. (Also as for liquids, it follows that a similar derivation for the $\Delta < 0$ case leads to a change in the sign of the detected $I$-spin signal (SI, Section II)). However, while such $\gamma_S/\gamma_I$ polarization transfer can be achieved for some single crystal orientations, such maxima cannot be achieved for all of them. This can be understood when considering the unavoidable "off-resonance" $\phi$-tilt in Eq. (28). Extensions of these calculations to isotropic powdered samples (including numerical simulations involving up to six spins) show that a maximum polarization transfer $\cong \frac{1}{2}\gamma_S/\gamma_I$ may in general be achieved.

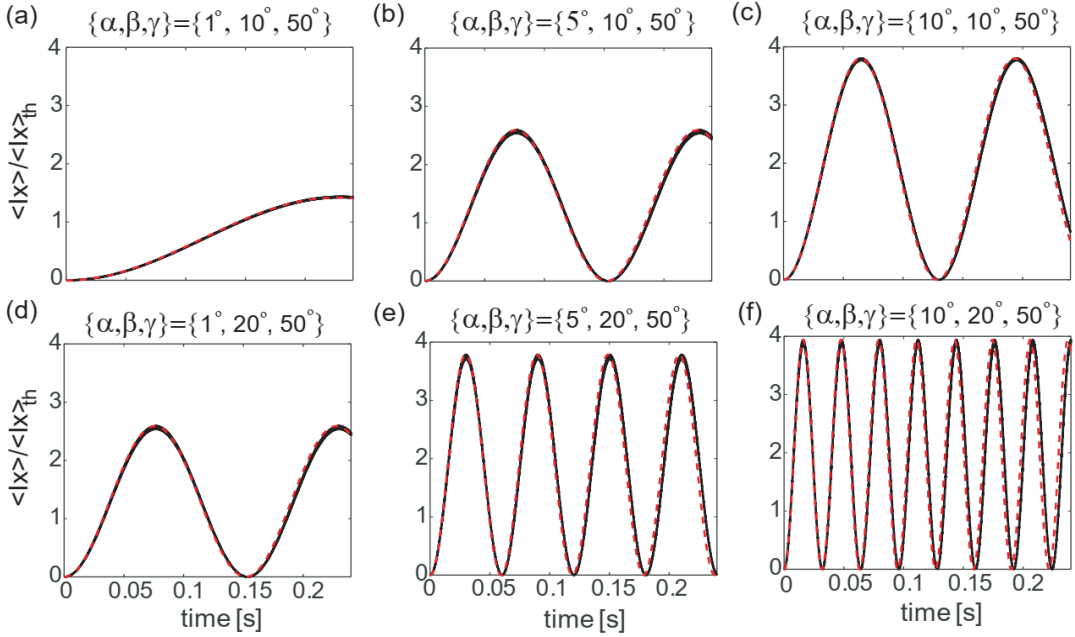

Figure 4. Comparison between the polarization transfer evolution predicted by Eq. (28) (red) and brute-force simulations (black), for the transfer of anti-longitudinal $H_{1z} - H_{2z}$ magnetization to $^{13}C_x$ transverse magnetization as a function of spin-lock time. In all cases the three nuclei were assumed arranged in an 'L' configuration, with $r_{H1H2} = 2.2$ Å, $r_{H2C} = 1.1$ Å, a chemical shift separation $\Delta/2\pi = 1$ kHz, MAS at 40 kHz, no chemical shift anisotropies, and an on-resonance $\omega_1/2\pi = 1$ kHz RF irradiation. Thermal $^1$H and $^{13}$C polarizations were set to 4 and 1, respectively. The crystallite orientations are characterized by the noted $\alpha$, $\beta$, $\gamma$ Euler angles with respect to a rotor fixed frame.

Figure 5 explores these multispin CP transfers further using numerical simulations involving five protons and one carbon, investigating their behaviour in powdered solids as function of spinning rate, RF-offset, and $\Delta$ (or Larmor frequency). The initial state considered was always $\rho(0) = -H_{1z} - H_{2z} + H_{3z} + H_{4z} + H_{5z}$, and the $^{13}$C spin was assumed bonded to protons 1 and 2. These numerical simulations recapitulate features from the analytical derivations above, including the fact that the isotropic $S$-spin chemical shift differences define the optimal CP transfer conditions, with these centered at $\omega_1 \cong \Delta$ and the matching being relatively broad thanks to the stronger spin-spin couplings. This ends up providing the 3-spin CP with a robustness vis-à-vis RF inhomogeneities that was absent in the liquids case. Especially at higher spinning rates, two matching conditions are observed. $\omega_1 \cong \Delta_1, \omega_1 \cong \Delta_2$ corresponding to the two possible chemical shift differences between the CH$_2$ protons and the other protons that were assumed. An increase in transfer efficiency is also observed at higher spinning rates, accompanied by a slight increase in the optimal time for maximizing the CP transfer. Also increasing the shift separation |Δ| leads in this case to a slight increase in the transfer efficiency. We ascribe these last two features to an "improvement" of the three-spin transfer process, whereby both faster MAS rates and larger chemical shift separations reduce the interferences from multispin phenomena that, as mentioned in the liquid state case, interfere with the efficiency of this kind of CP.

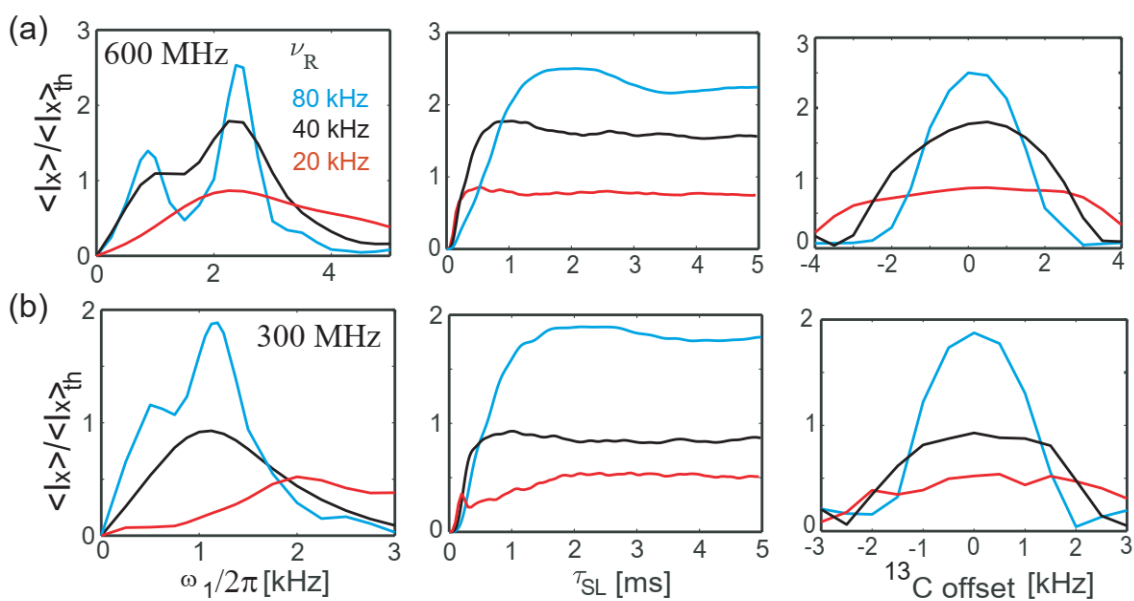

Figure 5. SIMPSON[26] powder simulations depicting polarization transfer to $^{13}C$ for three spinning rates, 20 kHz, 40 kHz and 80 kHz and five protons dipolar coupled to a $^{13}C$. (a) RF dependence of transfer (left) for $\Delta/2\pi = 2.5\ kHz$; time buildup (middle) for $\omega_1^{opt}/2\pi = 2.5\ kHz$ and three MAS rates; $^{13}C$ offset dependence of transfer (right) for $\nu_R = 40\ kHz$ and $\nu_o^{1H} = 600\ MHz$. (b) Similar to (a) but with $\Delta/2\pi = 1.25\ kHz$ corresponding to $\nu_o^{1H} = 300\ MHz$. The spin system contains 5 protons and a $^{13}C$ nucleus. The $^{13}C$ nucleus belongs to a $CH_2$ group containing protons 1 and 2. Protons 3, 4, and 5, have homonuclear dipolar couplings of a few kHz. The proton chemical shifts (at 600 MHz Larmor frequency), relative to the lowest chemical shift of proton 1, are $\nu(1) = 0\ Hz$, $\nu(2) = 126\ Hz$, $\nu(3) = 1008\ Hz$, $\nu(4) = 2538\ Hz$, $\nu(2) = 2538\ Hz$.

## 3) Materials and Methods

Solid State NMR experiments were performed using a Bruker Neo® console interfaced to an Oxford Instruments 14.1 T wide-bore magnet [$\nu_0(^1H)$ = 600 MHz], using a Varian/Chemagnetics 1.6 mm triple-resonance MAS probe. Spinning rates were regulated by a standalone Varian controller at 40 kHz. Solution state NMR experiments were performed on a Bruker 14.1 T magnet equipped with a TCI Prodigy® probe [$\nu_0(^1H)$ = 600 MHz] and on a Magnex 7 T [$\nu_0(^1H)$ = 300 MHz] widebore magnet interfaced to a Bruker console and QNP probe. XiX-12[27] with $\nu_{1H} = 20\ kHz$ proton decoupling was used in the solid $^{13}C$ acquisitions and WALTZ-16[28] in the liquid ones. All natural abundant samples here used –glycine, β-alanine, cystine, 4-nitrobenzonitrile– were purchased from Sigma-Aldrich and without further purification. Solids were packed as powders in 1.6 mm rotors (Revolution NMR), while liquid studies were done in 50 mM CDCl$_3$ solutions. All data were processed either using home-written Matlab® programs or Topspin 4.0. Simulations were carried out using home written Matlab® programs or the SIMPSON simulation package, as mentioned.

## 4) Results
### 4.a $^1$H Anti-Longitudinal ⇔ $^{13}$C Spin-Locked Three-spin CP in Solution-State NMR

Figure 6 shows a series of $^1$H and $^{13}$C NMR experiments on a 4-nitrobenzonitrile solution, testing the aforementioned liquid-state predictions. Protons (1,1') and (2,2') are separated in this compound by Δ ≈ 283 Hz, J-coupled by $\omega_J^{HH}/2\pi$ ≈ 8.5 Hz among themselves, and by $\omega_J^{H1C}/2\pi$ ≈ 172 Hz, $\omega_J^{H2C}/2\pi$ ≈ 7 Hz to the bonded and non-bonded $^{13}$C. Figure 6b shows results arising upon using the pulse sequence in Figure 1a to test the transfer of anti-longitudinal $^1$H magnetization in this compound, to $^{13}$C spin-locked magnetization. To ensure a clean background a purge period entailing fifty 90° $^{13}$C pulses (13 μs) with alternating phases and interspersed with unequal crasher gradients, was used; this lead to null $^{13}$C signals in the absence of heteronuclear polarization transfers. This purge was followed by an IBURP-2 [29] selective inversion pulse applied on-resonance on either the (1,1') or (2,2') protons (made and optimized using Topspin's shape-tool, 23.4 ms long, $\gamma B_1^{max.}$= 214 Hz); the left-hand panel shows vertically magnified $^1$H spectra arising from such selective inversion, evidencing a clean manipulation of both the $^{12}$C-center and $^{13}$C-satellite $^1$H peaks associated to these sites. When these inversions were followed by a spin-locking pulse applied on-resonance at the chemical shift of either the $^{13}$C$_{1,1'}$ or $^{13}$C$_{2,2'}$ carbons, a clear (signed) CP transfer was observed in the ensuing $^{13}$C acquisitions (Fig. 6b, right). In all cases, the experimental RF transfer profiles (Fig. 6c, black trace) exhibits the narrow maximum at the optimal amplitude $\omega_1^{opt}/2\pi$ = 260 Hz, as predicted by Eq. (12). As predicted by theory, this profile can be broadened and made less dependent on RF inhomogeneities by ramping the $^{13}$C RF fields (Fig. 6c, green race). Also shown in Fig. 6c are the $^{13}$C magnetization build-up curves as a function of spin-lock time in the continuous-wave and ramped cases, reaching in both instances a long-lasting plateau. Notice that this is in contrast to the oscillatory buildup predicted by theory for the minimal three-spin system (Fig. 2a); on the basis of multi-spin simulations (see below), we attribute this experimental behaviour to the influence of the multiple $^1$H-$^1$H J-couplings in the molecule.

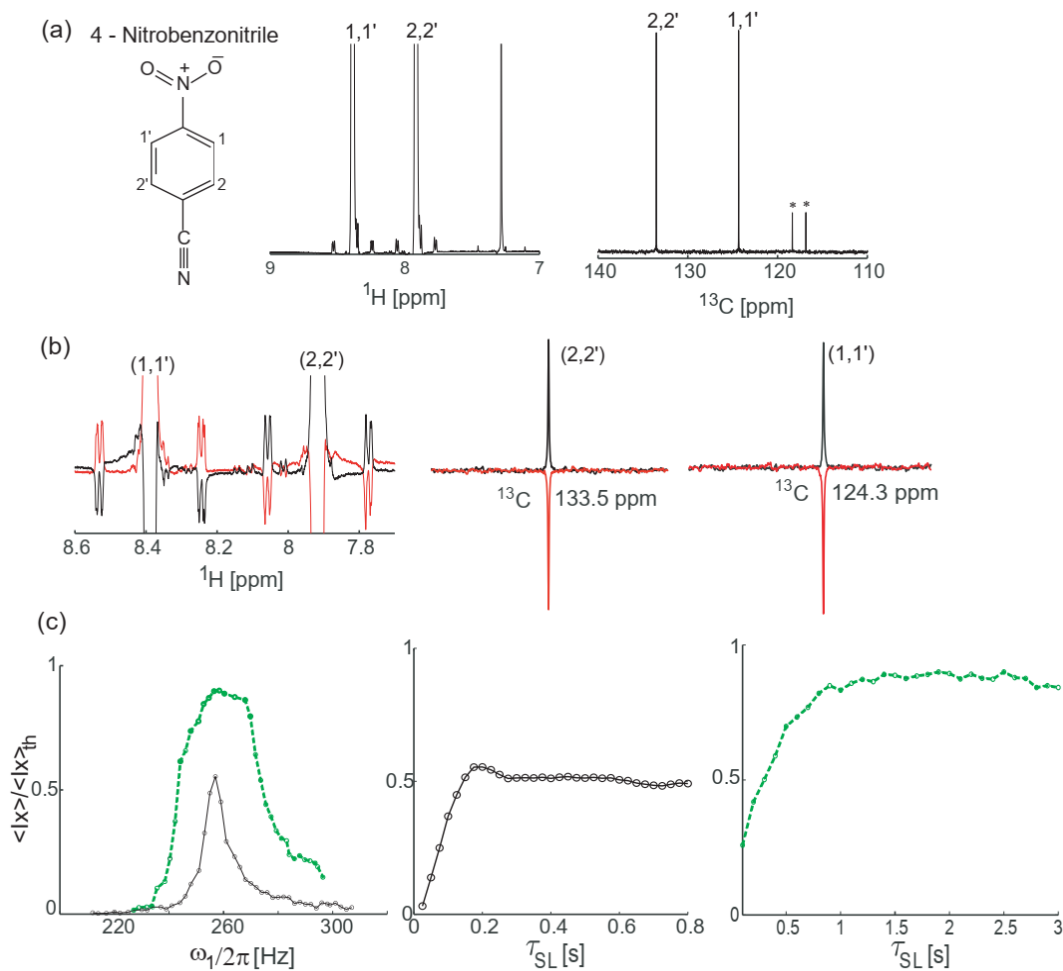

Figure 6. (a) 4-nitrobenzonitrile sample plus $^1$H and {$^1$H-decoupled}$^{13}$C 1D NMR spectra; the former showing the $^{13}$C satellites and the latter showing non-protonated carbons by asterisks. (b) Magnified $^1$H NMR spectrum showing the selective inversion of the satellites of (1,1', in black) and (2,2', in red), followed by an optimized CP three-spin transfer to $^{13}$C$_{1,1'}$ or $^{13}$C$_{2,2'}$, leading to the {$^1$H}$^{13}$C acquisitions on center and right. (c) RF mismatch of the CP process with upon using a continuous amplitude (black) vs a 5% ramped RF (green) for the $^{13}$C magnetization buildups shown in the center and right. All plots in (c) are normalized with respect to $\langle I_x \rangle_{th}$; notice in the center panel the beginning of an oscillatory behavior, that gets dampened by multispin effects (see text).

4-nitrobenzonitrile experiments were also performed at 300 MHz, where protons (1,1') and (2,2') are separated by $\Delta \approx 142$ Hz. Figure 7 summarizes the results observed on 4-nitrobenzene when repeated at this lower field. The optimum CP values are experimentally observed at $\omega_1^{opt}/2\pi = 112$ Hz, close to theoretical prediction, and for $\tau_{SL} = 900\ ms$ (Figure 7b). While at 600 MHz the continuous wave spin-lock resulted in about 50% loss in the $^{13}$C magnetization, these RF-inhomogeneity losses are smaller (~ 20%) at 300 MHz.

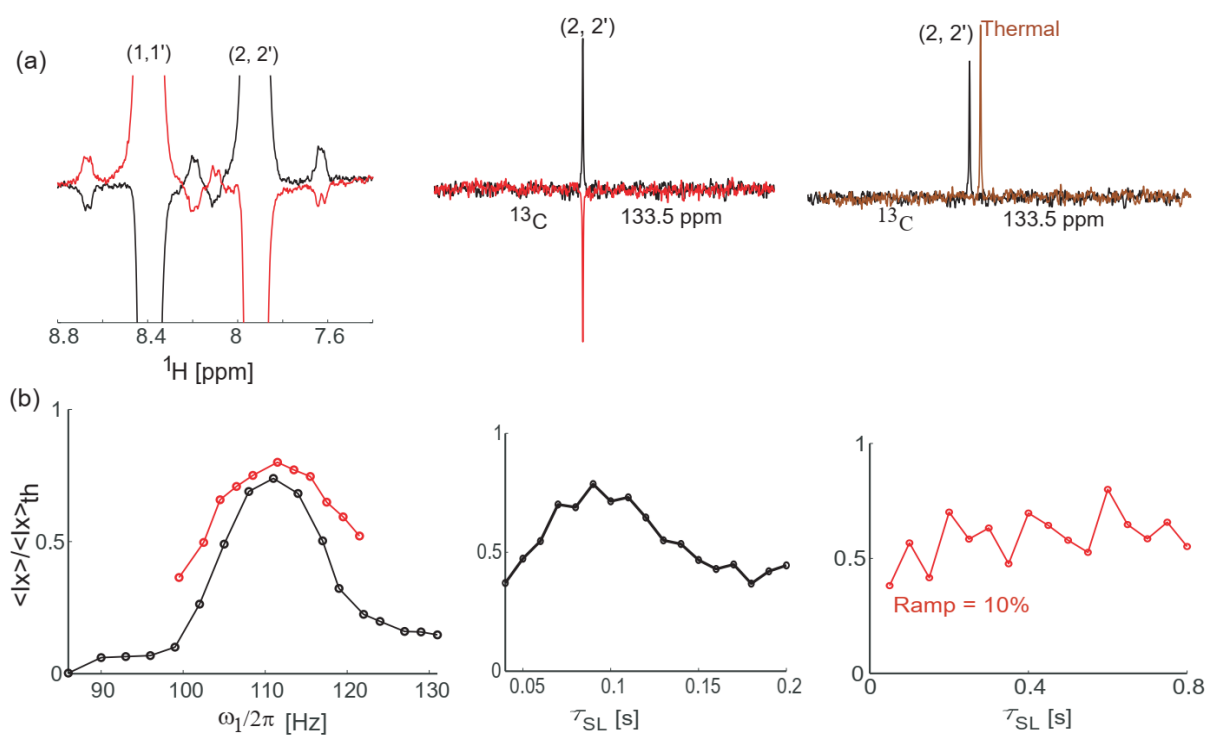

Figure 7. Summary of results obtained upon repeating the 3-spin CP experiments on 4-nitrobenzonitrile at 300 MHz; notice that because of the lower field, the $^{13}$C satellites of (1,1') and (2,2') protons cross over each other at ≈8.2 ppm (a, left). Selective inversions were achieved with a 27.4 ms long IBURP-2 pulse ($\gamma B_1^{max.}$= 181 Hz). Notice that the RF matching profiles are now centered at an optimum RF of about 112 Hz, and that ramps meant to compensate for RF inhomogeneities (in red) affect the transfers much less than at higher fields (b). The buildup provides an optimum transfer time of 90 ms with CW and no substantial sensitivity enhancement with longer transfer times with ramp. (see SI, Section V )

Further insight based on numerical simulations depicting the role of multispin interactions and RF inhomogeneity for 4-nitrobenzonitrile are shown in the SI, Section III. These suggest that achieving the full CP transfer is indeed hindered by multispin $^1$H-$^1$H interactions, which also compound the effects of RF heterogeneities. Another potential source of losses relates to potential $T_{1\rho}$ decays of the $^{13}$C magnetization during the long spin-locks involved. From the $T_{1\rho}$ decay shown in the SI (Section IV) it follows that these losses were < 10% for the ≈500 ms spin-lock times used, suggesting that at least for small molecules $T_{1\rho}$ relaxation is not primarily responsible for the sensitivity losses.

### 4.b $^1$H Anti-Longitudinal ⇔ $^{13}$C Three-spin CP in Solids under MAS

Similar NMR experiments were performed on powdered solid samples undergoing MAS. For testing these $^1$H→$^{13}$C transfers (Figure 1a), ten $^{13}$C 90° phase-cycled pulses (X/Y/-X/-Y… 4.5 µs each) separated by 1.5 ms delays were employed as initial purge, a Gaussian pulse (550 μs long, $\nu_1^{max}$ = 5.5 kHz) was used for the selective $^1$H inversion, and $^{13}$C transverse magnetization was detected with XiX-12 decoupling. Figure 8 summarizes the results obtained with these parameters for the case of powdered natural abundant glycine at 40 kHz MAS. The chemical shift difference between the NH$_3$ and CH$_2$ protons is here ~ 3 kHz; the chemical shift difference between the CH$_2$ protons is smaller (~600 Hz), and these were thus treated as equivalent. Inversion of either the CH$_2$ or the NH$_3$ protons (Fig. 8a, left panel) followed by an on-resonance (42 ppm) $^{13}$C spin-lock pulse, resulted in an observable $^{13}$C signal, whose phase reversed depending on the sign of the anti-longitudinal $^1$H magnetization (Fig. 8a, center). Figure 8b shows a search for the optimal $\tau_{SL}$ and $\omega_1^{opt}/2\pi$ needed to effect this heteronuclear CP; these were ≈2.85 ms and 3.6 kHz respectively. By contrast to the liquid-state case, this polarization transfer exhibits a rather broad matching around $\omega_1/2\pi \cong \Delta$, which can be attributed to the larger values and orientation dependence of the homonuclear dipolar interactions driving this process. The accumulation of $^{13}$C polarization with spin-lock time resembles what is observed under conventional HH CP; as can be seen from Fig. 7a (right-hand plots), the 3-spin CP also provides a similar sensitivity enhancement as that arising from ramped CPMAS under these conditions. However, contrary to what is observed under HH CPMAS and related CP sequences (rotary resonance, etc.) –for which the matching obeys $\omega_{1I} = \omega_{1S} \pm n\omega_R, n = 1,2$– the spin-lock RF amplitude will be here MAS-independent: it will only depend on the approximate chemical shift difference between the associated protons that participate in the anti-longitudinal state. This might become increasingly advantageous under very fast MAS, as the polarization transfer would then require only relatively small RF powers, and involve ever better resolved $^1$H sites. A potential conspiring factor could then be the spontaneous relaxation of the anti-longitudinal order, which might place a limit to the length of the spin-lock transfer pulse. This decay was quantified using the pulse sequence in Fig. 1a, but with a time delay incorporated between the selective inversion and the $^{13}$C spin-lock. As presented in the SI (Section VI) these anti-longitudinal proton states decay under MAS with relatively long lifetimes (≥100 ms), and do not act as limiting factor for these experiments.

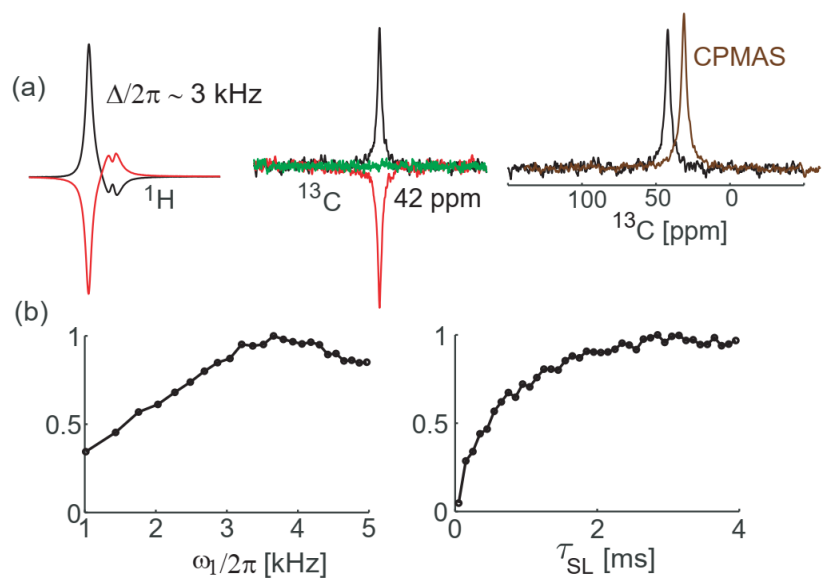

Figure 8. Summary of 3-spin CP results obtained on a glycine powder under 40 kHz MAS. (a) Selective inversion of the $CH_2$ protons (black) or the $NH_3$ protons (red), leading to opposite $^{13}C$ magnetizations for the $CH_2$ (~ 42 ppm) when recorded using the pulse sequence shown in Figure 1a and an on-resonance irradiation. Notice that no $^{13}C$ signal is observed without the selective inversion of the protons (green trace). The right-most panel shows that the ensuing $^{13}C$ signal (black) is comparable to that recorded using ramp-CP-MAS (brown) under optimal conditions (the latter spectrum is slightly shifted to the right to show a better comparison). (b) Additional details depicting the RF matching and the buildup curves leading to an optimum $^{13}C$ 3-spin CP enhancement.

Figure 9 complements these data with analogous experiments performed on other powdered solids. The overall behaviour is as that described for glycine. Figure 9a shows the experimental observations from β–alanine. Inversion of $NH_3$ (8.1 ppm) and $CH_2$ (1.3 ppm) protons separated by ~ 4 kHz along with a comparison with ramped CPMAS and polarization transfer from the anti-longitudinal proton spin states are shown (right). The corresponding RF match for optimum transfer (4 – 6 kHz), and the buildup curve (~ 3 ms) are also shown for the $^{13}CH_2$ peak when on-resonance (33.8 ppm). Figure 9b shows similar observation for cystine, which has two aliphatic carbons: a $^{13}CH_2$ (34 ppm) and a $^{13}CH$ (52.1 ppm). At 40 kHz MAS the protons of these groups are not resolved, leading to a spectrum with two peaks: $NH_3$ at 7.7 ppm, and $CH_2/CH$ at 2.9 ppm –ca. 3 kHz apart. Inverting the proton resonance at 2.9 ppm while placing the $^{13}C$ RF offset at either 34 or 52.1 ppm leads to enhanced $^{13}C$ signals at both shifts (red/magenta traces in Fig. 9b); likewise happens if inverting the proton resonance at 7.7 ppm (black). A comparison with ramped CPMAS spectrum (green) is also shown for both samples.

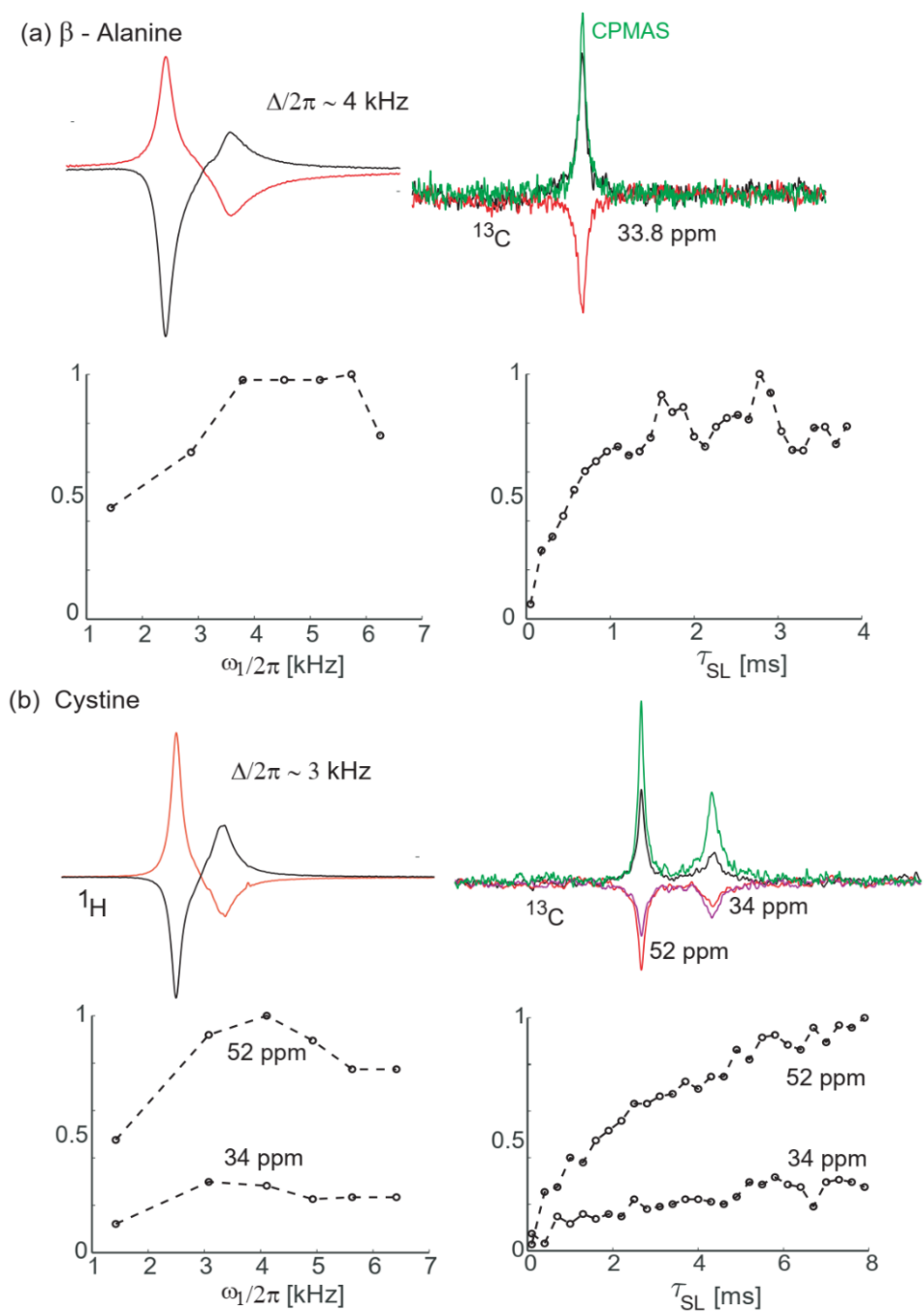

Figure 9: Idem as in Figure 8 but for: (a) β – alanine. (b) cystine. In both (a) and (b) optimized (ramped) HH CPMAS spectra collected at 40 kHz MAS, are also shown in green. For β–alanine the $^{13}$C spectra were recorded with $\omega_1^{opt}/2\pi$ =5.7 kHz and $\tau_{SL}$ = 1.9 ms. For the case of cystine, plots in red and magenta are shown for a $^{13}$C RF spin-lock set at 52 and 34 ppm, respectively. These data were recorded with $\omega_1^{opt}/2\pi$ = 4.2 kHz and $\tau_{SL}$ = 8 ms $^{13}$C RF fields. In the builup curves of cystine, even though at off-resonance, 34 ppm shows a similar behaviour. All plots are normalized with respect to maximal values.

Finally, Figure 10 exemplifies the reversed process, whereby an anti-longitudinal $^1$H state is created by spin-locking an excited $^{13}$C magnetization. To collect these results, about fifty 90° pulses of length 1.78 µs, were applied on the protons as purge, with 50 µs delays between them; no $^1$H spectrum could be observed thereafter. Then a $\tau_{SL}$ reversed spin-lock time was optimised for this reverse transfer (3.9 ms), and the same $\omega_{1C}$ was used as in Figure 8. Notice how an anti-longitudinal proton spin state then emerges under MAS, without ever pulsing on the $^1$Hs.

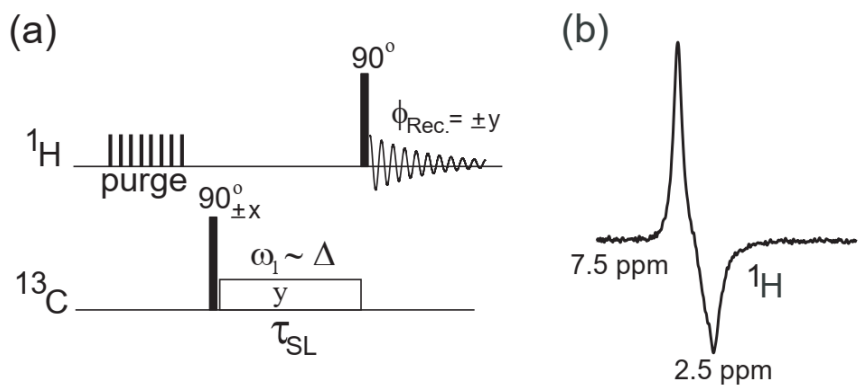

Figure 10: Inverse detection of anti-longitudinal proton magnetization using the pulse sequence in (a) on powdered glycine undergoing MAS at 40 kHz. A ±x phase cycling on the $^{13}$C 90° excitation pulse was used, and the $^1$H receiver is phase cycled as ±y. Additional experimental parameters are provided in the main text.

## 5) Discussion and Conclusions

This study explored and demonstrated a 3-spin CP transfer mechanism applicable to liquids and spinning solids, that starting from an anti-longitudinal magnetization state on a homonuclear coupled spin system, proceeds to create a spin-locked magnetization on a third, heteronuclear-coupled spin. Using a quantum mechanical treatment based on AHT it was found that, when executed under suitable matching conditions, this process is unitary; i.e., a spin-locked *I*-spin state can also become anti-longitudinal $S_{1z}$-$S_{2z}$ magnetization on a two-spin system. This, without ever having to pulse on the S-spins.

From the point of view of their dynamics, the evolution underlying these cyclic transfers end up resembling those arising under Hartmann-Hahn CP. However, by contrast to the first-order nature of HH CP, it is in all instances the non-secular two-spin coupling terms within the S-spin system, that enable the three-spin $S_1$–$S_2$⇔$I$ transfers to take place. AHT highlights this as non-zero second-order interferences arising between the homo- and heteronuclear interactions, and which under suitable conditions –an *I*-spin RF matching the $S_1$–$S_2$ chemical shift difference and thereby eliminating first-order effects– become the dominant factors in the ensuing spin evolution. To some extent this is reminiscent of the mechanism driving cross polarization from dipolar order: also in the ADRF-CP case it is a second-order interference between the homonuclear *S*-couplings and the heteronuclear coupling, that drives an oscillatory transfer from the dipolar reservoir to a spin-locked I-magnetization and back.[12] The experiments here introduced however, have a number of important differences that provide both advantageous and disadvantageous features vis-à-vis ADRF-CP. Counted among the advantages are (1) the fact that a starting anti-magnetization state can readily be made in isotropic liquids whereas an analogous dipolar state does not exist; and (2) the fact that dipolar order –including higher-order dipolar order states– are much less stable and harder to preserve upon fast MAS than anti-longitudinal states. In fact, for the case of spinning solids, sites become easier to resolve and more readily addressable in the $S_1$–$S_2 \Leftrightarrow I$ 3-spin CP here discussed, the faster the MAS rate becomes. This on the other hand also leads to a limitation of this method: the new CP form here described will not work well for protons at slow MAS, when chemically inequivalent sites will still remain unresolved. For the fast MAS cases no significant field dependence was revealed by simulations, but these were seen to affect the CP efficiencies in solutions, particularly when involving chemically inequivalent protons and high fields. The present study tried to bypass this by using relatively simple molecules as test cases, but even there challenges arose in the form of a high sensitivity to RF inhomogeneities, the need for long contact times, and a loss of sensitivity when extended $J_{HH}$ couplings competed against a pure 3-spin CP process. Preliminary experiments and simulations suggest that these complex $S$-spin networks do not influence strongly the efficiency of the 3-spin CP process in solids; in fact they become less of a complication the higher the field and the faster the MAS rate. This can make 3-spin CP a particularly attractive complement to other low-power CP methods in rapidly rotating solids.[31,32] Also the possibility of extending these experiments to higher spin numbers remains to be explored. These and other potential derivations of these novel form of cross-polarization, are under investigation.

**Acknowledgements**: This work was supported by ISF grant 1874/22, ERC Advanced Grant 101200719 "SteadyNMR", the Minerva Foundation, and the Perlman Family Foundation. SJ acknowledges the Weizmann Institute of Science for a Visiting Faculty fellowship. LF is the incumbent of The Bertha and Isadore Gudelsky Professorial Chair.

# SUPPORTING INFORMATION FOR

# Heteronuclear Polarization Transfer Between Spin-locked and Anti-longitudinal Spin States in the NMR of Liquids and Spinning Solids

Sundaresan Jayanthi,* Adonis Lupulescu,* Julia Grinshtein and Lucio Frydman*

Department of Chemical and Biological Physics, Weizmann Institute of Science, Rehovot, Israel

**Section I:** **Evaluating the Three-Spin Average Hamiltonian in Liquids**

The time dependent Hamiltonian below

$$\widetilde{\mathcal{H}}^{2-3}(t) = (\omega_1 - \Delta)I_z + \omega_J^{SS}\left[S_x^{2-3}\cos(\Delta t) - S_y^{2-3}\sin(\Delta t)\right]$$
$$- J_- S_z^{2-3}\left[I_x \cos(\Delta t) - I_y \sin(\Delta t)\right]. \qquad (S1.1)$$

can be represented as

$$\widetilde{\mathcal{H}}^{2-3}(t) = (\omega_1 - \Delta)I_z$$
$$+ \omega_J^{SS}\left[\frac{S_x^{2-3}}{2}(\exp(i\Delta t) + \exp(-i\Delta t)) - \frac{S_y^{2-3}}{2i}(\exp(i\Delta t) - \exp(-i\Delta t))\right]$$
$$- J_- S_z^{2-3}\left[\frac{I_x}{2}(\exp(i\Delta t) + \exp(-i\Delta t))\right.$$
$$\left. - \frac{I_y}{2i}(\exp(i\Delta t) - \exp(-i\Delta t))\right] \qquad (S1.2)$$

The first-order Hamiltonian is $\mathcal{H}_{ave}^{(1)} = \widetilde{H}_0$, and the second-order average Hamiltonian can be computed as

$$\mathcal{H}_{ave}^{(2)} = -\frac{1}{2}\sum_{p\neq 0}\frac{[\widetilde{H}_{-p},\widetilde{H}_p]}{p\Delta} + \sum_{p\neq 0}\frac{[\widetilde{H}_0,\widetilde{H}_p]}{p\Delta} \qquad (S1.3)$$

**Numerical Simulations with additional spins**. The effect of more $^1$H spins, RF-inhomogeneity, compensation with a linear ramp, and the effect of $B_o$ are shown in Figure S1. Numerical simulations showing the behaviour of $^{13}$C magnetization with the addition of one more proton (an $S_1 S_2 S_3 \, I$) in two different $B_o$ fields, with and without inhomogeneity, removal of inhomogeneity effects by the application of RAMP are shown below.

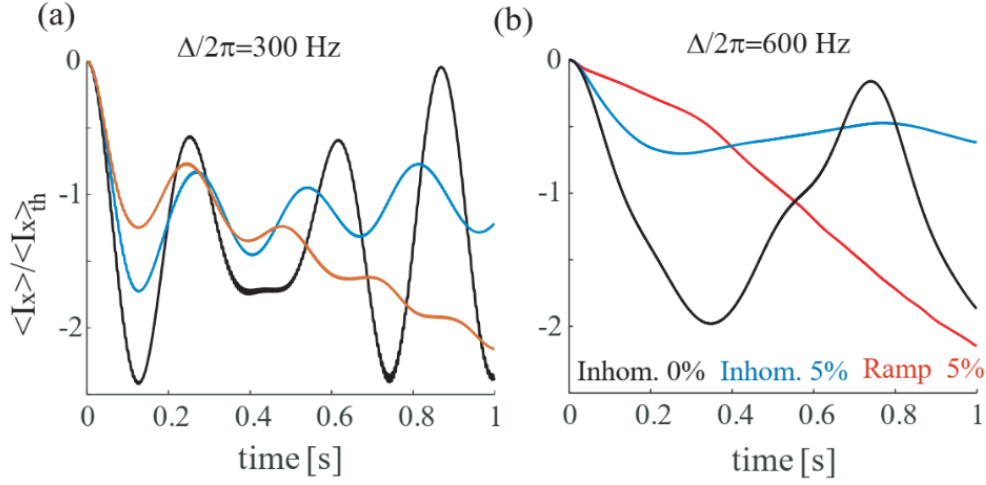

Figure S1. Numerical simulations depicting polarization transfer for the I spins for a $CH2H3 - {\bf 13}CH1$ spin system, with parameters, $J(H1 - H2) = 7\ Hz$; $J(H1 - H3) = 4\ Hz, J(13C - H1) = 170\ Hz$; $J(13C - H2) = J(13C - H3) = 8\ Hz$ where the chemical shift difference between $H1 - H2 = H1 - H3 = 200\ (400)\ Hz$ corresponding to a static field with proton resonance frequency at (a) 300 MHz and (b) 600 MHz

## Section II: Evaluating the Three-spin Average Hamiltonian in Spinning Solids

The tilted frame Hamiltonian is,

$$\mathcal{H}_T(t) = \omega_1 I_z + \Omega_1 S_{1z} + \Omega_2 S_{2z} + d_{SS}(t)[2S_{1z}S_{sz} - 1/2(S_{1+}S_{2-} + S_{1-}S_{2+})] - d_{S1I}(t)2S_{1z}I_x - d_{S2I}(t)2S_{2z}I_x \quad (S2.1)$$

We pass to an interaction frame following the transformation corresponding to Eq. [21] in the main text to get,

$$\widetilde{\mathcal{H}}^{2-3}(t) = -d_{SS}(t)\left[S_x^{2-3}\cos(\Delta t) - S_y^{2-3}\sin(\Delta t)\right] - 2D_-(t)S_z^{2-3}\left[I_x\cos(\Delta t) - I_y\sin(\Delta t)\right].$$

which can be written as,

$$\widetilde{\mathcal{H}}^{2-3}(t) = -d_{SS}(t)\left[\frac{S_x^{2-3}}{2}(\exp(i\Delta t) + \exp(-i\Delta t)) - \frac{S_y^{2-3}}{2i}(\exp(i\Delta t) - \exp(-i\Delta t))\right]$$

$$- 2D_-(t)S_z^{2-3}\left[\frac{I_x}{2}(\exp(i\Delta t) + \exp(-i\Delta t)) - \frac{I_y}{2i}(\exp(i\Delta t) - \exp(-i\Delta t))\right] \quad (S2.2)$$

$$\widetilde{\mathcal{H}}^{2-3}(t) = \sum_p \widetilde{H}_p \exp(ip\Delta t) \quad (S2.3)$$

The Fourier components $\tilde{H}_p$, where $p = nk \pm 1; n = \pm 1, \pm 2$. For example

$$\tilde{h}_{k+1} = -d_{ss}^{(1)} \left[\frac{1}{2}S_x^{2-3} + \frac{i}{2}S_y^{2-3}\right] - 2D_-^{(1)}S_z^{2-3}\left[\frac{1}{2}I_x + \frac{i}{2}I_y\right] = -\frac{1}{2}d_{ss}^{(1)}S_+^{2-3} - D_-^{(1)}S_z^{2-3}I_+.$$

$$\tilde{h}_{k-1} = -d_{ss}^{(1)} \left[\frac{1}{2}S_x^{2-3} - \frac{i}{2}S_y^{2-3}\right] - 2D_-^{(1)}S_z^{2-3}\left[\frac{1}{2}I_x - \frac{i}{2}I_y\right] = -\frac{1}{2}d_{ss}^{(1)}S_-^{2-3} - D_-^{(1)}S_z^{2-3}I_-.$$

The first order Hamiltonian $\mathcal{H}_{ave}^{(1)} = \tilde{h}_0$ is zero, hence we compute the second order average Hamiltonian as

$$\mathcal{H}_{ave}^{(2)} = -\frac{1}{2}\sum_{p\neq 0}\frac{[\tilde{h}_{-p}, \tilde{h}_p]}{p\Delta} + \sum_{p\neq 0}\frac{[\tilde{h}_0, \tilde{h}_p]}{p\Delta} = -\frac{1}{2}\sum_{p\neq 0}\frac{[\tilde{h}_{-p}, \tilde{h}_p]}{p\Delta}$$

$\mathcal{H}_{ave}^{(2)}$ has the following expression,

$$\mathcal{H}_{ave}^{(2)} = A(I_z + S_z^{2-3}) + B(I_z - S_z^{2-3}) + C_1 ZQ_x + C_2 ZQ_y,$$

with,

$A$

$$= \frac{\left\{d_{HH}^{(-1)}d_{HH}^{(+1)}(1-4k^2) + d_{HH}^{(-2)}d_{HH}^{(+2)}(1-k^2) + D_-^{(-1)}D_-^{(+1)}(1-4k^2) + D_-^{(-2)}D_-^{(+2)}(1-k^2)\right\}}{2\Delta(1-k^2)(1-4k^2)} \quad (S2.4)$$

$B$

$$= \frac{-\left\{d_{HH}^{(-1)}d_{HH}^{(+1)}(1-4k^2) + d_{HH}^{(-2)}d_{HH}^{(+2)}(1-k^2) - D_-^{(-1)}D_-^{(+1)}(1-4k^2) - D_-^{(-2)}D_-^{(+2)}(1-k^2)\right\}}{2\Delta(1-k^2)(1-4k^2)} \quad (S2.5)$$

$C_1$

$$= \frac{-\left\{D_-^{(-1)}d_{HH}^{(+1)}(1-4k^2) + D_-^{(+1)}d_{HH}^{(-1)}(1-4k^2) + D_-^{(-2)}d_{HH}^{(+2)}(1-k^2) + D_-^{(+2)}d_{HH}^{(-2)}(1-k^2)\right\}}{\Delta(1-k^2)(1-4k^2)} \quad (S2.6)$$

$C_2$

$$= \frac{-ik\left\{D_-^{(-1)}d_{HH}^{(+1)}(1-4k^2) - D_-^{(+1)}d_{HH}^{(-1)}(1-4k^2) + 2D_-^{(-2)}d_{HH}^{(+2)}(1-k^2) - 2D_-^{(+2)}d_{HH}^{(-2)}(1-k^2)\right\}}{\Delta(1-k^2)(1-4k^2)}$$

$$(S2.7)$$

**Calculations for Δ < 0.** For $\Delta < 0$ a similar approach leads to

$$\mathcal{H}_{ave}^{(2)} = A'(I_z + S_z^{2-3}) + B'(I_z - S_z^{2-3}) + C_1' DQ_x + C_2' DQ_y \qquad (S2.8)$$

where the new coefficients are linked to those defined in Eq. [S2.4 - 7] by $A' = -B$, $B' = -A$, $C_1' = C_1$, $C_2' = C_2$. It is now $A'(I_z + S_z^{2-3}) + C_1' DQ_x + C_2' DQ_y$ term which mediates the $S_1 S_2 \rightarrow I$ transfer. It is found that the final state is $\rho(t) = -\frac{1}{2}\sin^2\phi\left[1 - \cos(\omega_{eff} t)\right]I_x + \cdots$, where $\omega_{eff}$ is the same as before. A change in the sign of the detected *I*-spin signal is thus predicted with respect to the $\Delta > 0$ case.

## Section III: Role of multi-spin interactions and RF inhomogeneity in the liquid-state CP transfer

To further evaluate the effects of multispin interactions, the three-spin CP system described in the main text was extended to a five-spin system like the one present in 4-nitrobenzonitrile at natural abundance, encompassing protons (1,1') and (2,2') with mutual J-couplings, and a single $^{13}$C. The matching condition still remains nearly unchanged, yet the efficiency of the transfer drops rapidly below $\gamma_H/\gamma_C$ by the action of RF heterogeneities.

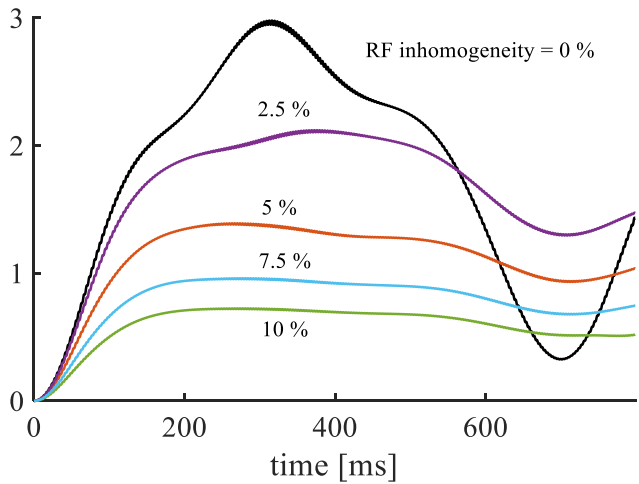

Figure S2. Simulations of an optimized $^1$H-$^{13}$C polarization transfer with parameters appropriate to 4-nitrobenzonitrile, for different degrees of RF inhomogeneity. With no inhomogeneity, the maximum achievable polarization is less than the theoretical expectation: A value of 3 is reached instead of 4. The buildup features upon introducing RF inhomogeneity suffer more than in the simpler three-spin system, and when about 5% they resemble the experimental buildup (Fig.5e). The parameters used for simulations are $\Delta/2\pi = 284\, Hz$, $J(H1 - H2) = J(H1' - H2') = 8.5 Hz$; $J(H1 - H1') = J(H2 - H2') = 2.5 Hz$; $J(H1 - C) = 172 Hz$; $J(H2 - C) = 7 Hz$, $\omega_1/2\pi = 266 Hz$.

## Section IV: $T_{1\rho}$ measurements of the $^{13}$C transverse magnetization

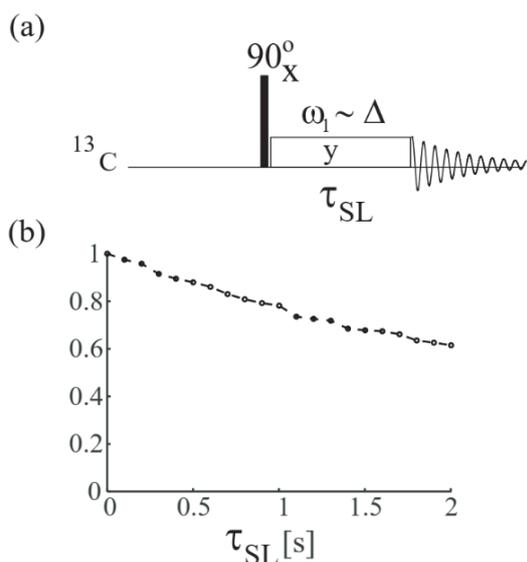

$T_{1\rho}$ measurement were carried out to estimate the percentage loss in $^{13}$C magnetization transferred in the liquid state, as a function of spin-lock time. This was performed to estimate the loss in the transferred $^{13}$C polarization due to relaxation during spin-lock. The ensuing results are illustrated in Figure S3.

Figure S3. (a) Pulse sequence utilized for measurement of $T_{1\rho}$. An RF power providing a 260 Hz Rabi nutation –corresponding to the CP matching conditions– was used, and a CW spin-lock was employed. (b) Normalized decay of $^{13}$C magnetization of the protonated 4-nitrobenzonitrile $^{13}$C during spin-lock. For $\tau_{SL}$=500 ms, the loss in magnetization is less than 10%.

## Section V: Role of multispin interactions and RF inhomogeneity at 300 MHz

Numerical simulations are carried out in 4-nitrobenzonitrile at natural abundance, encompassing protons (1,1') and (2,2') with mutual J-couplings, and a single $^{13}$C, as in section III, but the chemical shift difference of the protons (1,1') and (2,2') are reduced by half. The numerical simulations of CW buildup resemble Figure 7 in the text. Here RF inhomogeneity has minimal effect due to small Δ. Simulations with 10% ramp also shows negligible enhancement.

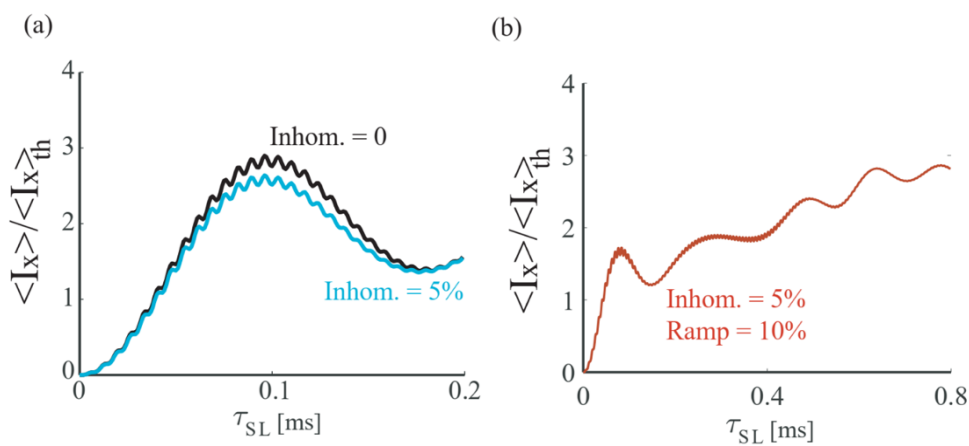

Figure S4. (a) Numerical simulations for 4-nitrobenzonitrile with a CW spinlock and Δ/2π ≈ 142 Hz, without and with 5% RF inhomogeneity. (b) With the CW spin-lock is replaced by 10% ramp.

# Section VI: Relaxation of anti-longitudinal proton spin-states in the solid state

The longitudinal relaxation of the anti-longitudinal proton spin-states are measured indirectly using the pulse sequence shown below, either by inverting the CH$_2$ protons or NH$_3$ protons in Glycine spinning at 40 kHz MAS. Experimental observation is fitted with a mono-exponential decay and the relaxation rates with root mean square error is shown in the respective plots. The T$_1$ relaxation of CH$_2$ protons differs drastically when compared with NH$_3$ protons as shown below.

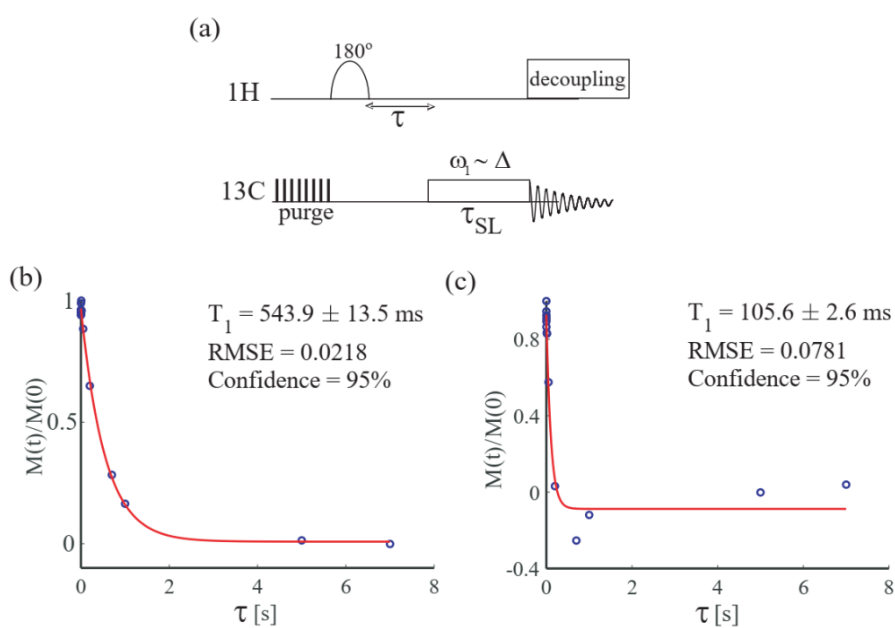

Figure S5. (a) Pulse sequence employed to measure the longitudinal relaxation rate of the anti-longitudinal protons states through indirect detection on $^{13}$C. (b) Results obtained with CH$_2$ inverted. (c) Idem with NH$_3$ inverted. In both cases a mono exponential decay function of the form $M(t) = M(0)exp(-\tau/T_1) + C$ was used to get the fit (red). A discrepance can be see in (c) whereby the $^{13}$C magnetization traverse through positive to zero, reverses, and eventually decays to zero. This oscillation was repeatedly observed, but not incorporated this behaviour in the fit.